\documentclass[twocolumn,aps,prl]{revtex4-1}

\usepackage[T1]{fontenc}

\usepackage{graphicx}
\usepackage{amsmath,amssymb}
\usepackage{bm}
\usepackage{textcomp,gensymb,units}
\usepackage[usenames]{color}

\newcommand{\vc}[1]{\ensuremath{\bm{#1}}}
\newcommand{\abs}[1]{\ensuremath{| {#1} |}}
\newcommand{\avg}[1]{\ensuremath{\langle {#1} \rangle}}
\newcommand{\var}[1]{\ensuremath{\Delta{#1^2}}}

\newcommand{\unitz}{\ensuremath{\vc{\hat z}}}

\newcommand{\ket}[1]{\ensuremath{|{#1}\rangle}}

\newcommand{\Sz}{\ensuremath{S_z}}
\newcommand{\Sperp}{\ensuremath{S_\perp}}

\newcommand{\Smeanlength}{\ensuremath{\abs{\avg{\vc{S}}}}}
\newcommand{\So}{\ensuremath{S_0}}

\newcommand{\allandev}{\ensuremath{\sigma}}
\newcommand{\sqclock}{\ensuremath{\zeta}}

\newcommand{\Rb}{\ensuremath{{}^{87}\mathrm{Rb}}}
\newcommand{\Fone}{\ket{F=1,m_F=0}}
\newcommand{\Ftwo}{\ket{F=2,m_F=0}}

\newcommand{\Dtwo}{\ensuremath{\mathrm{D}_2}}
\newcommand{\Tcoh}{\ensuremath{T_\text{coh}}}
\newcommand{\Tramsey}{\ensuremath{T_\text{R}}}
\newcommand{\Contrast}{\ensuremath{C}}
\newcommand{\Cini}{\ensuremath{\Contrast_\text{in}}}

\begin{document}

\title{Orientation-Dependent Entanglement Lifetime in a Squeezed Atomic Clock}

\author{Ian D. Leroux}
\affiliation{Department of Physics,
  MIT-Harvard Center for Ultracold Atoms and Research Laboratory of
  Electronics, Massachusetts Institute of Technology, Cambridge,
  Massachusetts 02139, USA}

\author{Monika H. Schleier-Smith}
\affiliation{Department of Physics,
  MIT-Harvard Center for Ultracold Atoms and Research Laboratory of
  Electronics, Massachusetts Institute of Technology, Cambridge,
  Massachusetts 02139, USA}

\author{Vladan Vuleti\'{c}}
\affiliation{Department of Physics,
  MIT-Harvard Center for Ultracold Atoms and Research Laboratory of
  Electronics, Massachusetts Institute of Technology, Cambridge,
  Massachusetts 02139, USA}

\date{\today}

\begin{abstract}
  We study experimentally the lifetime of a special class of entangled
  states in an atomic clock, squeezed spin states.  In the presence of
  anisotropic noise, their lifetime is strongly dependent on squeezing
  orientation.  We measure the Allan deviation spectrum of a clock
  operated with a phase-squeezed input state.  For integration times
  up to \unit[50]{s} the squeezed clock achieves a given precision
  2.8(3) times faster than a clock operating at the standard quantum
  limit.
\end{abstract}

\maketitle

Atomic interference provides an exquisitely sensitive tool for
measuring gravitation, magnetic fields, acceleration, rotation, and
time itself~\cite{Cronin09,Vanier89}.  It has long been hoped that
quantum-mechanical entanglement might enhance the precision of such
measurements: maximally-entangled states can increase the sensitivity
of the interference fringe to the parameter of
interest~\cite{Bollinger96}, while squeezed spin states can
redistribute quantum noise away from that
quantity~\cite{Wineland92,Wineland94}.  In experiments, both
approaches have overcome the standard quantum limit (SQL) of phase
sensitivity~\cite{Meyer01,Leibfried04,Schleier-Smith08,Appel09,Leroux10,Louchet-Chauvet10}.
However, Huelga et al.\ pointed out early on that entangled states
might provide little gain in real metrological performance because
they are more fragile than uncorrelated states, such that the
entanglement-induced increase in phase sensitivity comes at the
expense of reduced interrogation time~\cite{Huelga97}.  Analyses with
specific noise models~\cite{Wineland98,Andre04}, however, found
parameter regimes where entanglement could be helpful despite
decoherence.  It is thus interesting, practically as well as
fundamentally, to investigate the fragility of the entangled states
relevant to metrology.

In this Letter we show that for an atomic clock in which the dominant
environmental perturbation is phase noise, the squeezed-state lifetime
varies by an order of magnitude depending on whether the squeezed
variable is the phase (subject to environmental perturbation) or the
(essentially unperturbed) population difference between states.  We
operate an atomic clock with a phase-squeezed input state whose
precision exceeds the SQL, as also recently demonstrated by
Louchet-Chauvet et al.~\cite{Louchet-Chauvet10}, and present the first
measurement of such a clock's Allan deviation spectrum.  The clock
reaches a given precision 2.8(3) times faster than the SQL for
integration times up to \unit[50]{s}.  The squeezed states used in
this work are prepared by cavity feedback
squeezing~\cite{Schleier-Smith10,Leroux10}, a new technique which
deterministically produces entangled states of distant atoms using
their collective interaction with a driven optical resonator.

\begin{figure}
  \centering
  \includegraphics[width=\columnwidth]{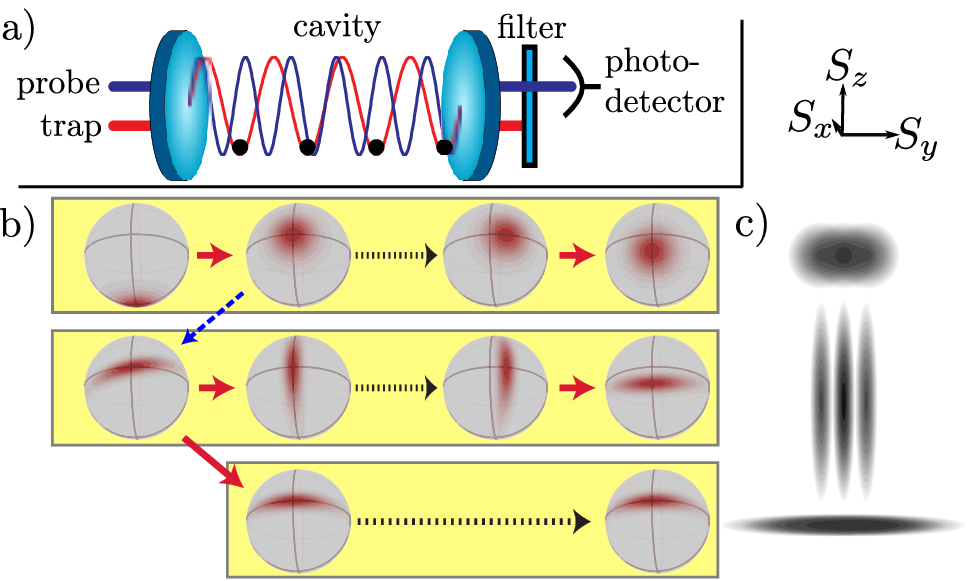}
  \caption{\mbox{\textbf{(a) Setup:}} A standing-wave dipole potential
    confines \Rb\ atoms inside an optical resonator where they
    interact with probe light via their \Sz-dependent index of
    refraction.  The probe light is used for cavity feedback
    squeezing~\cite{Schleier-Smith10,Leroux10} and final-state
    readout.  \mbox{\textbf{(b) Pulse Sequences:}} A standard Ramsey
    protocol (top bar) consists of optical pumping into the spin-down
    CSS, rotation into the equatorial plane using a microwave pulse
    (red arrow), free precession time (dotted black arrow), and
    conversion of the accrued phase into a population difference for
    readout.  We can shear the CSS into a squeezed
    state using cavity feedback (dashed blue arrow), then orient the
    narrow axis of the squeezed state in the phase direction (middle
    bar) or in the \unitz\ direction (bottom bar).  \mbox{\textbf{(c)
        Effect of Phase Noise:}} At the end of the free precession
    time, classical uncertainty on the accrued phase broadens the CSS
    in one direction (top).  The same noise is detrimental to
    phase-squeezed states (middle) but much less noticeable in
    number-squeezed states (bottom) whose narrow direction is not
    affected.}
  \label{fig:setup}
\end{figure}

Given any two-level atom we can define a spin-$1/2$ $\vc{s}_i$.  For
an ensemble of such atoms, we introduce the total spin
$\vc{S}=\sum\vc{s}_i$ whose \unitz\ component and azimuthal angle
$\phi$ represent the population difference and relative phase,
respectively, between the two atomic levels.  Simultaneously preparing
the atoms in the same single-particle quantum state places the
ensemble in a coherent spin state (CSS) where the variance of the spin
components perpendicular to the mean spin is given by
$\Smeanlength/2$~\cite{Kitagawa93}.  While loss of coherence between
the atoms can shorten $\Smeanlength$ and reduce this variance, only
entanglement between atoms can improve the signal-to-noise ratio (SNR)
of a measurement of the orientation of
$\vc{S}$~\cite{Wineland92,Wineland94}.  For a system with $2\So$ total
atoms and signal contrast \Contrast\ (\Cini) for the correlated
(uncorrelated) state after (before) the squeezing procedure, we can
define a metrological squeezing parameter~\cite{Wineland92} $\sqclock
= 2 \var{\Sperp} \Cini / (\So\Contrast^2)$, which compares the squared
SNR for the best possible measurement on a CSS with the spin length
$\Cini\So$ to that of the actual measurement with transverse variance
\var{\Sperp} and spin length $\Smeanlength = \Contrast\So$.  If
$\sqclock < 1$, the total spin orientation is more precisely
determined, in some direction, than would be possible if the atoms
were not entangled.  The spin state is then squeezed, and the
individual atoms in the ensemble are necessarily
entangled~\cite{Sorensen01}.

If we prepare a CSS in the equatorial plane, we expect that the phase
noise will increase at long times due to classical fluctuations on the
energy difference between levels in our apparatus
(Fig.~\ref{fig:setup}c, top).  As discussed below, this broadening
becomes noticeable in our system in less than $\unit[1]{ms}$.  In the
number direction, the primary mechanism that increases polar angle
uncertainty $\Delta\Sz/\Smeanlength$ is loss of contrast (reduction in
the length of the mean spin vector $\abs{\avg{\vc{S}}}$), which occurs
on a longer time scale $\Tcoh=\unit[11(1)]{ms}$ in our apparatus.  A
phase-squeezed state will therefore suffer much more rapid broadening
of its narrow axis (Fig.~\ref{fig:setup}c, middle) than a
number-squeezed state (Fig.~\ref{fig:setup}c, bottom).  Intriguingly,
a number-squeezed state is less vulnerable to phase noise than an
uncorrelated state: an increase in phase variance by several times the
width of the original CSS might still be small compared to the
antisqueezed phase variance of the number-squeezed state.  Finding
robust orientations for the squeezed state is reminiscent of the
search for decoherence-free subspaces in quantum information
science~\cite{Kwiat00,Kielpinski01,Viola01}: with knowledge of the
specific noise processes in a particular system, one can find quantum
states which are relatively immune to their effects.

We work with laser-cooled \Rb\ and use the canonical
magnetic-field-insensitive clock transition, choosing \Fone\ and
\Ftwo\ as the two clock states.  The atomic cloud is held by a dipole
trap inside a Fabry-P\'{e}rot resonator (Fig.~\ref{fig:setup}a), one
mode of which is detuned halfway between the \Dtwo\ optical
transitions for the two clock states.  The total spin $\vc{S}$
corresponds to a sum over the atomic cloud, weighted by the atoms'
position-dependent coupling to the resonator mode so as to yield an
effective uniform-coupling description~\cite{Leroux10}.  The index of
refraction of the atoms shifts the cavity resonance frequency by equal
and opposite amounts for atoms in each of the two clock states, the
net shift being proportional to their population difference $2\Sz$.
In order to read out the atomic state, the resonator is driven by
probe light tuned to the slope of the cavity resonance, so that
atom-induced shifts of the resonance frequency are revealed as changes
in the transmitted fraction of probe light, which we detect on an
avalanche photodiode.  The probe light is generated as a sideband of a
laser locked to another cavity mode, \unit[33]{GHz} red-detuned from
the \Dtwo\ transition for the $F=2$ state in order to reduce the
interaction between lock light and atoms.  A \unit[2.4]{G} bias field
along the cavity axis combined with circular trap polarization makes
the clock frequency first-order independent of trap
power~\cite{Schleier-Smith08}, at the price of making it linearly
sensitive to magnetic field fluctuations with a coefficient of
$\unit[3.7]{kHz/G}$.  These fluctuations are the dominant noise
affecting the clock spin.  Aside from the choice of lock detuning,
bias field, and trap polarization given above, the details of our
apparatus and method for preparing squeezed spin states follow the
description of Refs.~\cite{Schleier-Smith08,Leroux10}.

Arbitrary rotations of the spin vector can be performed using resonant
microwave pulses, but squeezing the uncertainty region requires an
effective interaction between the atoms which we generate by cavity
feedback, exploiting their common coupling to the light field of the
resonator~\cite{Schleier-Smith10,Leroux10}.  As the atomic index of
refraction shifts the cavity resonance by an amount proportional to
\Sz, it changes the intracavity intensity of probe light.  Since the
probe imparts a light shift to the atoms, each atom acquires a phase
shift which depends on the state of all other atoms in the ensemble,
thus introducing the correlations necessary for squeezing.  The
\Sz-dependent phase shift shears the circular uncertainty region of
the CSS into an ellipse with its long (antisqueezed) axis oriented at
a small and known angle to the equatorial plane
(Fig.~\ref{fig:setup}b, middle bar).  Note that, due to photon shot
noise in the probe light, the states thus prepared are actually mixed
states with an area much larger than is required by the Heisenberg
uncertainty relations~\cite{Schleier-Smith10}.

\begin{figure}
  \centering
  \includegraphics[width=\columnwidth]{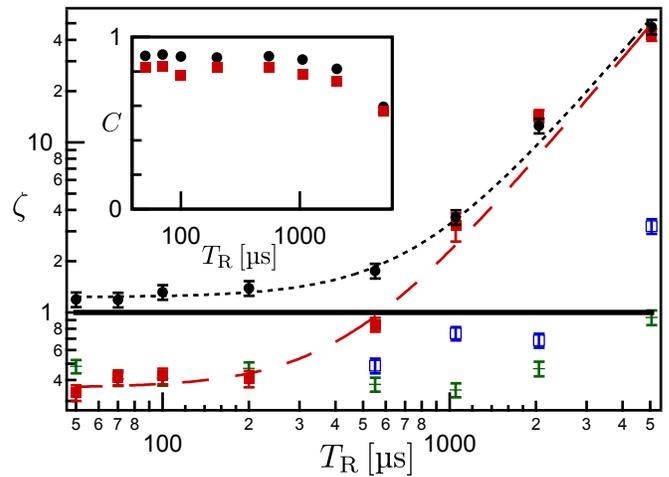}
  \caption{Metrological squeezing parameter \sqclock\ as a function of
    time.  The phase variance for an initial CSS (black circles,
    dotted line) or phase-squeezed state (red solid squares, dashed
    line) increases due to classical frequency noise.  The number
    variance of a number-squeezed state (green dashes) or the phase
    variance of a phase-squeezed state protected by a spin echo (blue
    open squares) can remain below the projection noise limit (solid
    black line) substantially longer.  Error bars show the statistical
    uncertainty of the variance determination.  Inset: Signal contrast
    \Contrast\ for the CSS (black circles) and squeezed states (red
    squares).}
  \label{fig:lifetime}
\end{figure}

To measure the lifetime of a phase-squeezed state, we load the dipole
trap with an ensemble of atoms collected in a magneto-optical trap
(MOT), prepare a CSS by optically pumping the atoms into the \Fone\
state and then apply a microwave $\pi/2$ pulse to rotate it into the
equatorial plane of the Bloch sphere~\cite{Leroux10}.  We typically
work with $2\So\approx 3\times10^4$ effective atoms and an initial
contrast $\Cini=90(2)\%$, yielding a projection-noise-limited phase
uncertainty of $\sim\!\unit[6]{mrad}$~\cite{Wineland92}.  Cavity
feedback squeezing with two weak pulses of probe light then drives the
atoms into a state with $\sqclock^{-1}\approx\unit[4]{dB}$.  A
rotation of nearly $\pi/2$ converts this into a phase-squeezed state.
After allowing the spins to precess for a variable time \Tramsey, the
phase information is converted back to a population difference with a
final $\pi/2$ pulse and read out by observing the transmission of a
pair of strong probe pulses.  This sequence of operations constitutes
a Ramsey-type atomic clock with a squeezed input state.  We perform 10
sequences of state preparation, precession and readout for each sample
of atoms loaded from the MOT, and the entire experimental cycle
repeats every 9 seconds.  We measure the phase variance of the CSS
using the same experimental sequence but without the probe light which
performs the squeezing.  To measure the lifetime of a number-squeezed
state, we rotate the sheared state slightly to orient its narrow axis
along \unitz\ and then simply hold it for a time \Tramsey\ before
reading it out (Fig.~\ref{fig:setup}b, bottom bar).  In all three
cases, comparing the normalized variance to the squared contrast for
the readout signal yields the metrological squeezing parameter
\sqclock, which we plot in Fig.~\ref{fig:lifetime}.  The SQL (black
line at $\sqclock=1$) is calculated from independent absolute atom
number measurements based on precisely-measured cavity parameters, and
verified experimentally as in Ref.~\cite{Schleier-Smith08}.

We begin by using the CSS to evaluate the classical phase noise
(Fig.~\ref{fig:lifetime}, black circles).  The data are well-described
by the model
$\sqclock(\Tramsey)=\sqclock(0)+2\So\Cini\var{\omega}\Tramsey^2$
(dotted fit), involving an initial angular uncertainty described by
$\sqclock(0)$ and additional fluctuations of the transition frequency
between measurements with variance $\var{\omega}$.  After a precession
time of $\approx\!\unit[700]{\micro s}$, the effect of the classical
noise $\Delta\omega = 2\pi\times\unit[1.3]{Hz}$ exceeds the initial
projection noise and the phase variance increases quadratically
thereafter.  Note that reaching the SQL in an atomic clock requires
not only projection-noise-limited state preparation and readout, but
also an interrogation time short enough that quantum projection noise
remains the dominant uncertainty on the clock signal; less than
\unit[700]{\micro s} in this case.

When we prepare a phase-squeezed state (solid red squares), we
initially observe a reduced phase variance $\sqclock(0)<1$, as
expected.  The same frequency noise $\var{\omega}$ broadens the
phase-squeezed state somewhat sooner than the CSS because there is
less quantum noise to mask the classical fluctuations.  Note, however, that
the same frequency noise that destroys the squeezed states in
$\approx\!\unit[600]{\micro s}$ also degrades the performance of
initially unsqueezed states, so that states that were initially
squeezed provide slightly better SNR even after $\sqclock(t)$
increases beyond unity.

Matters are substantially different when we prepare a number-squeezed
state and read out its reduced \unitz-variance directly after a hold
time \Tramsey\ (Fig.~\ref{fig:lifetime}, green dashes).  In this case
we are not operating a clock, which measures the evolution of the
phase angle, but instead examining the evolution of the polar angle
corresponding to the population difference between the clock states.
Frequency noise does not add uncertainty to this spin component, and
so \sqclock\ can remain below unity (squeezed) for \unit[5]{ms}, eight
times longer than for the phase-squeezed state, until dephasing
between the atoms, visible as loss of signal contrast
(Fig.~\ref{fig:lifetime}, inset), creates a mixed state whose quantum
correlations are insufficient to overcome the SQL.  There is no
measurable change in the phase variance of the number-squeezed state
(not shown) out to \unit[5]{ms}; the classical frequency noise is
entirely hidden in the antisqueezed initial phase variance.

Since the decay of the phase-squeezed state results from classical
phase noise, standard techniques can suppress it.  For example, we
have operated a ``clock'' sequence with a phase-squeezed input state,
but with an additional spin-echo pulse halfway through the precession
time.  The final phase is then insensitive to the atomic transition
frequency, which protects the state from slow frequency fluctuations
but makes it useless for timekeeping.  The state remains squeezed
\unit[2]{ms} after being prepared in the otherwise fragile
phase-squeezed orientation (Fig.~\ref{fig:lifetime}, open blue
squares).

Note that several experimentally demonstrated spin squeezing
schemes~\cite{Esteve08,Schleier-Smith08,Appel09,Leroux10} prepare the
ensemble in or near the robust number-squeezed state.  It is therefore
possible to prepare squeezed input states for an atomic clock at
leisure, while remaining largely insensitive to perturbations of the
atomic phase.  It is only when the clock is started by a subsequent
$\pi/2$ pulse \cite{Louchet-Chauvet10} that the squeezed state becomes
sensitive to frequency noise.

\begin{figure}
  \centering
  \includegraphics[width=\columnwidth]{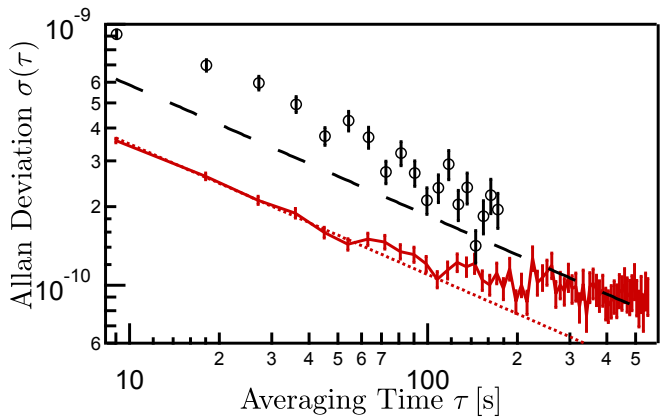}
  \caption{Allan deviation of a squeezed clock.  The solid red line
    with error bars was measured using a squeezed input state.  The
    dotted red line indicates
    $\allandev(\tau)=\unit[1.1\times10^{-9}]{s^{1/2}}/\sqrt{\tau}$.
    The open black circles were measured with a traditional clock
    using an uncorrelated input state.  The dashed black line at
    $\unit[1.85\times10^{-9}]{s^{1/2}}/\sqrt{\tau}$ indicates the SQL
    at 100\% signal contrast.}
  \label{fig:allandev}
\end{figure}

As a demonstration, we have operated such a clock with a Ramsey
interrogation time $\Tramsey=\unit[200]{\micro s}$, short enough that
the classical frequency noise in our system does not destroy the phase
squeezing.  The effective atom number was $2\So=3.5\times10^4$, the
clock cycle time was \unit[9]{s} and the signal contrast was $C=81\%$.
Only a single Ramsey interrogation was performed for each MOT loading
cycle, giving a duty factor of $2\times10^{-5}$.  The result is the
first measurement of Allan deviation~\cite{Vanier89} for an atomic
clock operating beyond the SQL, including all noise and slow drifts
(Fig.~\ref{fig:allandev}, red solid line).  For comparison, we also
evaluate a clock operated with an uncorrelated input state close to a
CSS, 100(2)\% signal contrast and otherwise identical parameters
(Fig.~\ref{fig:allandev}, black circles).  An ideal
projection-noise-limited clock with the same atom number,
interrogation time and duty factor could reach a stability
$\allandev(\tau)=\unit[1.85\times10^{-9}]{s^{1/2}}/\sqrt{\tau}$
(Fig.~\ref{fig:allandev}, dashed black line).  At short times our
squeezed-state clock reaches a fractional frequency uncertainty of
$\allandev(\tau)=\unit[1.1\times10^{-9}]{s^{1/2}}/\sqrt{\tau}$, a
factor of 2.8(3) in variance below the SQL~\cite{Santarelli99}.  At
longer times we reach a noise floor at $10^{-10}$ fractional stability
(\unit[0.7]{Hz} absolute stability) due to slow drifts of the magnetic
field in our apparatus.

The performance of our clock only benefits from squeezing because we
impose the constraint of a short Ramsey precession time.  For longer
interrogation times the classical noise dominates the initial phase
noise and our clock is not projection-noise-limited.  There are,
however, realistic scenarios where external constraints or local
oscillator dephasing limit the allowable Ramsey precession time and
where metrological performance could be improved by the use of
squeezed input states~\cite{Wineland98,Andre04}.  For clocks with duty
factors near unity~\cite{Lodewyck09}, the greater sensitivity of a
squeezed clock also enables the investigation of systematic effects in
a shorter integration time.

If the classical frequency noise could be controlled at the level of
$\sim\!\unit[100]{\micro Hz}$, perhaps by operating in a magnetic trap
at the magic bias field of \unit[3.23]{G}, the squeezed lifetime could
be extended sufficiently to allow a squeezed clock to operate with a
Ramsey precession time of one second, as demonstrated by Treutlein et
al.\ on a microchip similar to the one used in the present
experiment~\cite{Treutlein04}.  Even without improvements to our
squeezed-state preparation or \unit[9]{s} cycle time, this could yield
a short-term stability of
$\allandev(\tau)\approx\unit[2\times10^{-13}]{s^{1/2}}/\sqrt{\tau}$,
competitive with the stability of current fountain
clocks~\cite{Gerginov10}.

This work was supported in part by the NSF, DARPA, and the NSF Center
for Ultracold Atoms.  M.~H.~S. acknowledges support from the Hertz
Foundation and NSF.  I.~D.~L. acknowledges support from NSERC.

%
\end{document}